# Visualization and Manipulation of Bilayer Graphene Quantum Dots with Broken Rotational Symmetry and Nontrivial Topology

Zhehao Ge[1,*], Fredric Joucken[1,*], Eberth A. Quezada[1,*], Diego R. da Costa[2], John L. Davenport[1], Brian Giraldo[3], Takashi Taniguchi[4], Kenji Watanabe[5], Nobuhiko P. Kobayashi[3], Tony Low[6], Jairo Velasco Jr.[1,†]

[1]*Department of Physics, University of California, Santa Cruz, California 95064, USA*

[2]*Departamento de Física, Universidade Federal do Ceará, Caixa Postal 6030, Campus do Pici, 60455-900 Fortaleza, Ceará, Brazil*

[3]*Jack Baskin School of Engineering, University of California, Santa Cruz, California 95064, USA*

[4] *International Center for Materials Nanoarchitectronics National Institute for Materials Science, 1-1 Namiki, Tsukuba, 305-0044, Japan*

[5] *Research Center for Functional Materials National Institute for Materials Science, 1-1 Namiki, Tsukuba, 305-0044, Japan*

[6]*Department of Electrical and Computer Engineering, University of Minnesota, Minneapolis, Minnesota 55455, USA*

[*]These authors contributed equally to this manuscript.
[†]Email: jvelasc5@ucsc.edu

**Abstract:**

Electrostatically defined quantum dots (QDs) in Bernal stacked bilayer graphene (BLG) are a promising quantum information platform because of their long spin decoherence times, high sample quality, and tunability. Importantly, the shape of QD states determine the electron energy spectrum, the interactions between electrons, and the coupling of electrons to their environment, all of which are relevant for quantum information processing. Despite its importance, the shape of BLG QD states remains experimentally unexamined. Here we report direct visualization of BLG QD states by using a scanning tunneling microscope. Strikingly, we find these states exhibit a robust broken rotational symmetry. By using a numerical tight-binding model, we determine that the observed broken rotational symmetry can be attributed to low energy anisotropic bands. We then compare confined holes and electrons and demonstrate the influence of BLG's nontrivial band topology. Our study distinguishes BLG QDs from prior QD platforms with trivial band topology.

KEYWORDS: Quantum dots, Bilayer graphene, Berry Curvature, Quantum Information, STM

The visualization and manipulation of electronic states in quantum materials is of fundamental interest and has the potential for quantum information processing technologies[1, 2]. Among the numerous quantum material platforms that are under consideration for these advanced technologies Bernal stacked bilayer graphene (BLG) is highly attractive because it possesses an electric field tunable band gap[3]. This unique material property permits the realization of electrostatic quantum dots (QDs) with pristine boundaries and high flexibility to enable charge carrier confinement and QD state formation[4, 5]. In addition, BLG QDs have long spin decoherence lifetimes[6], controllable quantum degrees of freedom[7, 8], and nontrivial band topology[9]. These traits are all favorable for quantum information technology. Recent experimental progress with BLG QDs interfaced with sheets of hexagonal boron nitride (hBN) have demonstrated important advancements regarding quantum state control such as spin-valley resolved single electron charging[7, 10] and magnetic field controlled valley splitting[7, 8]. Of equal importance, is the understanding of the QD wave function shape, which determines the energy spectrum and allowable transitions between states. These energetics govern the preparation, manipulation, and readout of quantum information[11]. Theoretical predictions for BLG QD wave function shape are abundant[12-17]; however, the experimental validation of these predictions is lacking.

Here we use a scanning tunneling microscope (STM) tip to create an exposed circular p-n junction in a BLG/hBN heterostructure for imaging and manipulating new QD states. To create such nanostructures, we employed a tip voltage pulsing technique based on prior works with the formation of circular p-n junctions in graphene/hBN heterostructures[18, 19] (see Supplementary Section 1 for more information on tip pulsing technique). A schematized circular p-n junction and measurement setup is depicted in the left panel of Fig. 1a. Our STM tip is grounded and a voltage bias ($V_S$) is applied to the sample. A gate voltage ($V_G$) is also applied to the underlying doped Si and is used to tune the BLG Fermi level.

To avoid the influence of adsorbates in our studies we create circular p-n junctions in $200 \times 200$ nm² adsorbate free regions. The middle panel of Fig. 1a shows the topography of one such region. The furthest right panel of Fig. 1a shows an atomically resolved topography at the center of the region where a circular BLG p-n junction was created. Bright topographical features that form a triangular lattice are clearly visible, as expected for BLG[20, 21]. We overlay the atomic structure of BLG in this panel and use it to identify three of the four atoms within the BLG unit cell: $A_{top}$, $B_{top}$, and $B_{bottom}$. The fourth atom ($A_{bottom}$) on the bottom layer is obscured by $B_{top}$. In such a triangular lattice two orientations exist for the apparent triangles, we denote these orientations as $\alpha$ (dashed blue line that encloses $B_{top}$) and $\beta$ (dashed yellow line that encloses $B_{bottom}$).

By performing constant sample bias $dI/dV_S$ maps we image the electronic states inside and outside of our circular BLG p-n junctions. In Figs. 1b and 1c we show maps taken at two different energies $V_s = -10.5$ mV and $V_s = -12.5$ mV, but at the same $V_G = -5$ V. A dashed yellow line provides a guide to the eye for the p-n junction. The electronic states outside of the p-n junction exhibit an approximate circular symmetry consistent with the p-n junction shape itself. In contrast, the states within the p-n junction display a rich nodal pattern that breaks rotational symmetry. For example, in Fig. 1b there is a "Y" shaped feature that can be seen at the center of the p-n junction with three antinodes that surround it. The pattern in Fig. 1c, on the other hand, contains three distinct antinodes at the center of the junction. Despite the diverse nature of the patterns in each map, a clear $C_3$ symmetry is shared between both of them. Closer inspection of these electronic states also reveals that they are aligned with the orientation of the $\alpha$ triangle of the BLG lattice (see Supplementary Section 4 for additional data on $C_3$ symmetry and BLG lattice alignment reproducibility).

To attain a more comprehensive understanding of the spatial distribution for the electronic states within our BLG p-n junctions we acquired $dI/dV_S$ spectra along zigzag and armchair directions across the center of the p-n junction from Fig 1b and 1c. The results of these measurements are shown in Figs. 2a and 2b. The color scale corresponds to $dI/dV_S$ amplitude and is plotted as a function of $V_S$ and the distance across the circular p-n junction. These measurements were taken at $V_G = -5$ V. A distinct nodal level structure with high $dI/dV_S$ amplitude is visible and is surrounded by a dark envelope (low $dI/dV_S$ amplitude) in each of the two plots. Moreover, for the $dI/dV_S$ spectra taken along the armchair direction (Fig. 2a), the nodal distribution exhibits an asymmetric $dI/dV_S$ intensity with respect to the center of the p-n junction. On the other hand, along the zigzag direction (Fig. 2b), the nodal distribution of the same p-n junction exhibits a clear mirror symmetry with respect to the center of the junction. The well-defined difference seen between the armchair and zigzag directions indicates that the broken rotational symmetry observed in Figs. 1b and 1c persists for a wide energy range.

Insight into the origin of our observations can be gained by considering the behavior of BLG charges corralled within a circular electrostatic potential well that is defined by a circular p-n junction. This setup has extensive previous theoretical inquiry, yet as we will show, our experimental findings cannot be explained by these prior works. The touching BLG bands at the p-n junction are opened by the presence of finite a perpendicular electric field[22, 23], thus yielding charge carrier confinement and formation of QD states within the interior region of the circular p-n junction[4, 5, 12, 24]. The dark envelope in Figs. 2a and 2b indicates a reduction of electronic states at the p-n junction, hence revealing the presence of an electric field induced gap. Consequently, the nodal pattern enclosed by this electric field induced gap corresponds to QD states in our experiment. The energetic and spatial distribution of these states are reminiscent of charges confined within a harmonic potential. For BLG QDs, most theoretical investigations predicted circularly symmetric states[12-16] akin to QD states seen in monolayer graphene[18, 25]. Notably, the $C_3$ symmetry that is clearly present in our QD states breaks rotational symmetry and are thus inconsistent with these prior predictions.

Interestingly, the electronic states within our QD exhibits the same symmetry as anisotropic low energy BLG bands that arise from the incorporation of a skewed interlayer hopping term $\gamma_3$[26]. This connection suggests an explanation for our findings and is also supported by a recent BLG QD theory which considered $\gamma_3$ hopping[17]. To verify this possibility, we proceeded by using numerical tight-binding (TB) calculations of a BLG QD that incorporates $\gamma_3$ hopping and a band gap. The potential well and the size of the band gap used in this calculation were extracted from the experiment (see Supplementary Section 5 for details on the numerical calculation and model[27]). Based on this calculation we simulated $dI/dV_S$ spectra along BLG zigzag and armchair crystallographic directions that cross the center of the potential well. These simulations are shown in Fig. 2c for the armchair direction and Fig. 2d for the zigzag direction. Remarkably, the nodal distribution and dark envelope in the simulated $dI/dV_S$ spectra closely resemble the measured spectra. For instance, the mirror symmetry about the center of the QD is clearly evident (absent) for the zigzag (armchair) direction(s). This symmetric (asymmetric) distribution of QD states along the zigzag (armchair) direction(s) can be attributed to the preservation (breaking) of mirror symmetry for BLG QDs along the zigzag (armchair) direction(s). The agreement between simulation and experiment is more apparent by comparing constant energy $dI/dV_S$ profiles as shown in Figs. 2e and 2f. A strong agreement is also seen between experimental $dI/dV_S$ maps and simulated constant energy local density of states (LDOS) maps (See Supplementary Section 6 for comparison). Crucially, we found that this agreement breaks down if either $\gamma_3$ or a band gap is not included in the numerical TB calculation (See Supplementary Section 7 for comparison).

The observed strong influence of gapped and anisotropic bands for BLG QD states has intriguing implications for accessing the nontrivial band topology of BLG. This is because such bands lead to a finite Berry curvature that is also anisotropic[28]. To explore this possible attribute of our QD system we use simulations based on numerical TB calculations to study QDs with p-type and n-type central doping and with different interlayer potential difference polarities ($U > 0$ and $U < 0$). As shown in Fig. 3a, the different types of QDs with gapped and anisotropic bands enable trapping of low energy holes (upper panel) and low energy electrons (lower panel). Simulated LDOS maps for analogous QDs with the same potential well curvature and gap size but with different polarity for the interlayer potential difference are shown in Figs. 3b-e. First, we focus on QDs with p-type central doping (Figs. 3b and 3c). We find that the orientation of the LDOS distribution for these QDs exhibits a 180° rotation between $U = +60$ meV and $U = -60$ meV. In addition, for $U = +60$ meV, the LDOS pattern is aligned with the $\alpha$ triangle of the BLG lattice, while for $U = -60$ meV the LDOS pattern is aligned with the $\beta$ triangle of the BLG lattice. A similar rotation and LDOS pattern alignment with respect to the BLG lattice can also be seen for QDs with n-type central doping (Figs. 3d and 3e).

The rotation in the simulated LDOS patterns indicates that a property of BLG has been modified between the different interlayer potential difference polarity configurations. The band structure of BLG does not depend on the interlayer potential difference polarity (See Supplementary Section 9 for TB calculation of BLG band structure[26]). In contrast, the Berry curvature, which is related to the BLG band topology, does depend on the sign of the interlayer potential difference[28-30]. For example, for confined holes, the Berry curvature of BLG's valence band in the $K$ valley is negative for $U = +60$ meV (see inset of Fig. 3b). The Berry curvature of the same band is positive for $U = -60$ meV (see inset of Fig. 3c). A similar Berry curvature sign flip can be seen for QDs with n-type central doping as well (insets of Figs. 3d and 3e). Here we focus on one valley because the Berry curvature sign flip also occurs in the other valley when the interlayer potential polarity changes. Additionally, because time reversal symmetry is preserved in our BLG QDs, we expect that the two valleys remain degenerate and have the same LDOS distribution. The LDOS distribution can also be explained by considering that the Berry curvature sign and Fermi surface orientation are reversed in the two valleys (See Supplementary Section 10 for further discussion). Thus, the simulated LDOS patterns in Figs. 3b-3e suggest that imaging the states within BLG QDs can reveal Berry curvature manipulation.

To experimentally verify this intriguing possibility, we created a circular p-n junction at the same region as the previous junctions but with different central doping type. We used this approach because in our experiment we were unable to modify the polarity of the interlayer potential difference (See Supplementary Section 11 for additional details on interlayer potential difference[23]). Nonetheless, our simulations in Figs. 3b and 3d reveal a 180º rotation between trapped holes and electrons. In addition, the LDOS pattern for trapped holes (electrons) is aligned with the $\alpha$ ($\beta$) triangle of the BLG lattice. These predictions from the simulations are consistent with the Berry curvature sign flip for the corresponding bands shown in the insets of Figs. 3b and 3d[28-30]. Notably, Figs. 4a and 4b show the measured $dI/dV_S$ maps for QDs with p-type and n-type central doping, respectively. In exact agreement with the simulations, the orientations of the $dI/dV_S$ patterns differ by a 180° rotation and they align with the $\alpha$ triangle of the BLG lattice for trapped holes and with the $\beta$ triangle of the BLG lattice for trapped electrons.

In conclusion, we have spatially mapped BLG QD wave functions and demonstrated manipulation of their Berry curvature. We found that consideration of gapped and anisotropic low energy bands is crucial for understanding the behavior of electrostatically corralled BLG charges. Additionally, we revealed manifestations of the nontrivial BLG band topology by imaging the quantum interference in our BLG QDs. This finding could inspire future works using quantum interference to study the properties of materials that have finite Berry curvature such as semiconducting transition metal dichalcogenides[31] (e.g. $MoS_2$), topological insulators[32] (e.g. $WTe_2$) and Weyl semimetals[33-35] (e.g. TaAs). The technical advancements presented here can also be used to study more sophisticated platforms for quantum information technology such as BLG double and multiple QDs. The tunable couplings in these systems make them promising for the realization of quantum bits based on spin and valley degrees of freedom[36, 37].

**Acknowledgments:** We thank the Hummingbird Computational Cluster team at UC Santa Cruz for providing computational resources and support for the numerical TB calculations performed in this work.

**Author contributions:** J.V.J., Z.G., F.J., and A.Q. conceived the work and designed the research strategy. Z.G. and F.J. performed data analysis under J.V.J.'s supervision. Z.G., A.Q., J.D., and B.G. fabricated the samples under N.P.K. and J.V.J.'s supervision. K.W. and T.T. provided the hBN crystals. Z.G., F.J., and A.Q. carried out tunneling spectroscopy measurements under J.V.J.'s supervision. Z.G. performed numerical TB calculations and simulations with input from D.R.C. and T.L. D.R.C. performed additional calculations under T.L.'s supervision. Z.G. F.J. and J.V.J. developed interpretation for experimental findings with input from D.R.C. and T.L. J.V.J., Z.G., and F.J. wrote the paper. All authors discussed the paper and commented on the manuscript.

**Funding Sources:** J.V.J. acknowledges support from the National Science Foundation under award DMR-1753367 and the Army Research Office under contract W911NF-17-1-0473. K.W. and T.T. acknowledge support from the Elemental Strategy Initiative conducted by the MEXT, Japan, Grant Number JPMXP0112101001, JSPS KAKENHI Grant Numbers JP20H00354 and the CREST(JPMJCR15F3), JST.

# Figure 1

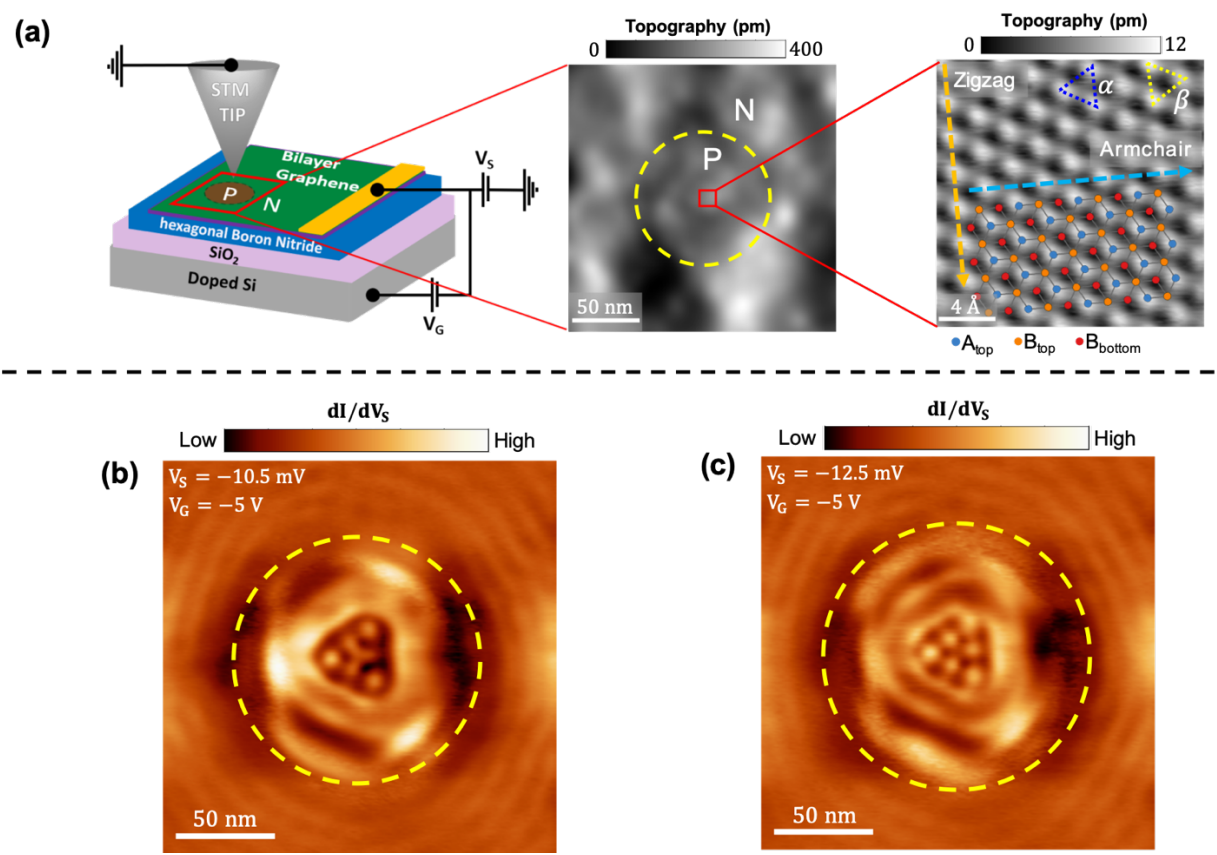

**Figure 1: Scanning tunneling microscopy/spectroscopy (STM/STS) characterization of a circular bilayer graphene p-n junction.** (a) **Left panel:** Schematic of a STM tip probing a circular p-n junction in Bernal stacked bilayer graphene (BLG) that is supported by hexagonal boron nitride (hBN). The BLG/hBN heterostructure rests on a SiO$_2$/doped Si wafer. The STM tip is grounded, a sample bias $V_S$ is applied between the BLG and ground, and a gate voltage $V_G$ is applied between the doped Si and BLG. **Middle panel:** BLG topography after the creation of a circular p-n junction. Dashed yellow line denotes the p-n junction. The scanning parameters were $V_S = -8.5$ mV, $I = 0.3$ nA. **Right panel:** Atomically resolved topography of BLG at the center of the p-n junction, the scanning parameters were $V_S = -100$ mV, $I = 2.5$ nA. BLG crystal structure is overlaid on the topography. The orange (blue) arrow indicates a BLG zigzag (armchair) direction(s). The blue and yellow triangles represent the two possible orientations for the BLG lattice by connecting the three closest bright atoms. **(b-c)** Constant sample bias $dI/dV_S$ maps of circular BLG p-n junctions at $V_G = -5$ V (b, c). The yellow dashed line in (b) and (c) denotes the boundary between p and n doped regions. Scanning parameters for (b) and (c) are $V_{ac} = 2$ mV and $I = 0.3$ nA.

**Figure 2**

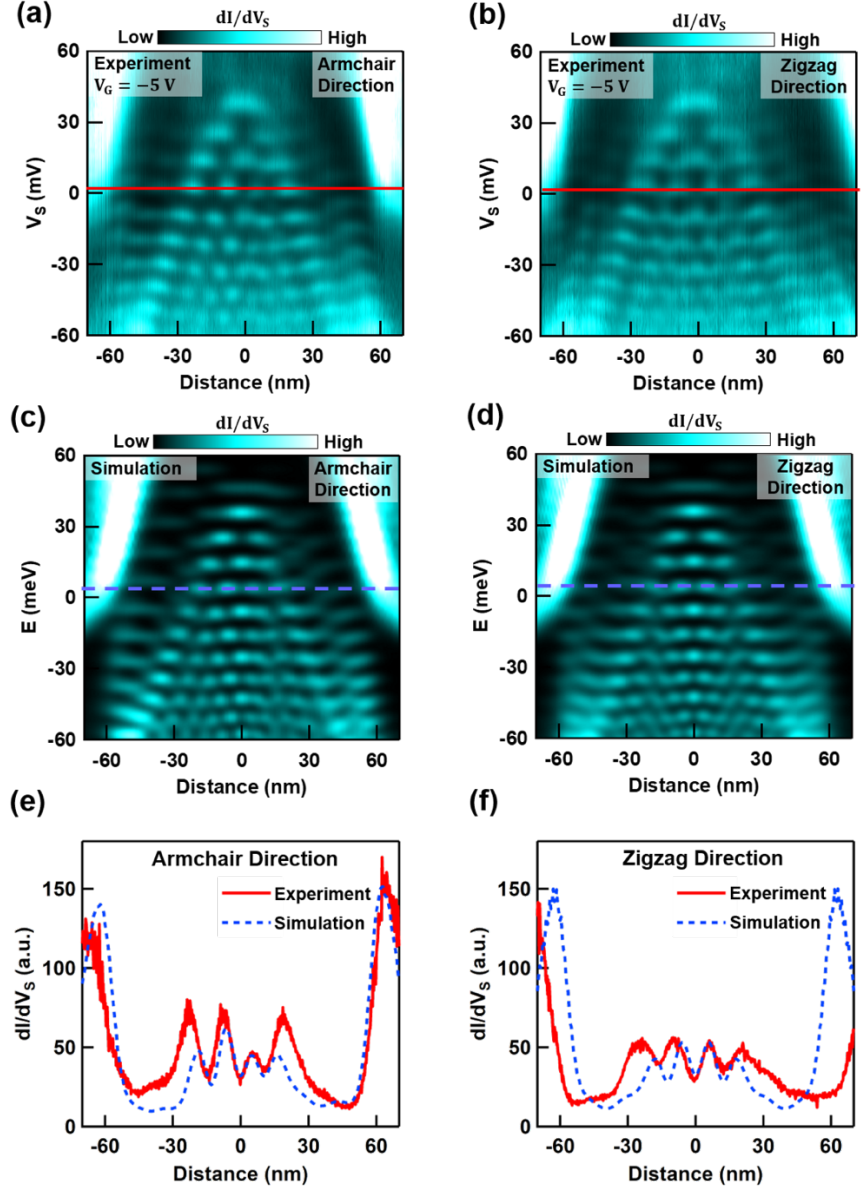

**Figure 2: Spatially resolved energy levels inside a BLG QD along zigzag and armchair directions. (a-b)** Measured $dI/dV_S$ spectra along BLG's armchair (**a**) and zigzag (**b**) directions at $V_G = -5$ V. The "0" on the horizontal axis in (a) and (b) denotes the center of the QD. The set point used in (a) and (b) was $I = 1$ nA, $V_S = -60$ mV, with a 2 mV ac modulation. **(c-d)** Simulated $dI/dV_S$ spectra along the BLG armchair and zigzag direction, respectively. The "0" on the horizontal axis in (c) and (d) denotes the center of the QD in the TB model. A constant 60 meV gap (top layer was set at a higher energy than the bottom layer, $U = +60$ meV) and $\gamma_3 = -0.38$ eV was included in the model. **(e)** Comparison between experimental and simulated constant energy $dI/dV_S$ profile along the BLG armchair direction. The experiment $dI/dV_S$ profile was acquired from (a) at $V_S = 2$ mV, indicated by the red solid line. The simulated $dI/dV_S$ profile was acquired from (c) at $E = 4$ meV, indicated by the blue dashed line. **(f)** Comparison between experiment and simulated constant energy $dI/dV_S$ profile along zigzag direction. Theses profiles were acquired at the same energies as in (e).

**Figure 3**

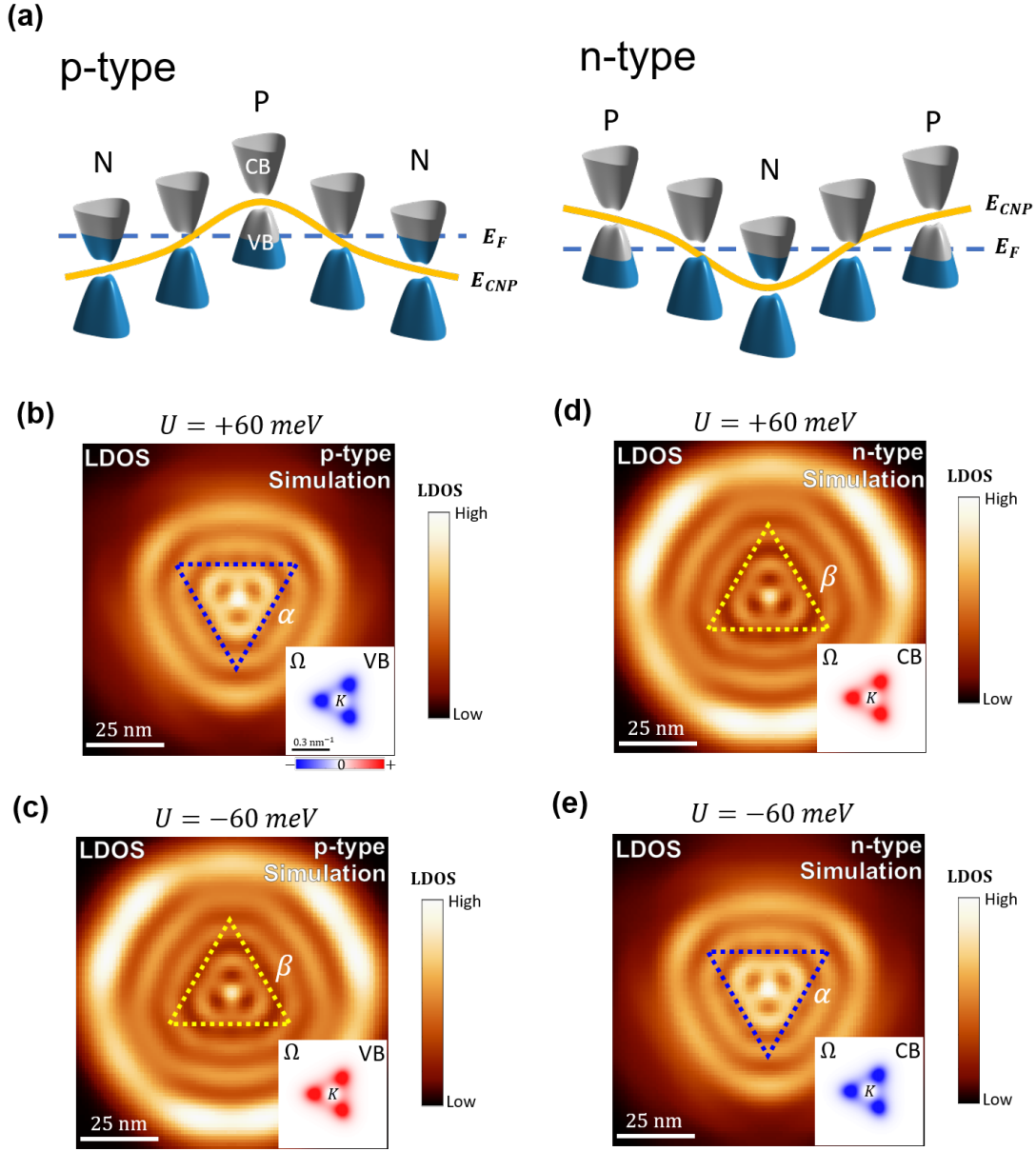

**Figure 3: Simualted BLG QD LDOS tracks the Berry curvature sign switch. (a)** Band diagrams for BLG QD with p-type (upper panel) and n-type (lower panel) central doping. The orange and blue curves indicate the BLG charge neutrality point ($E_{CNP}$) and Fermi level ($E_F$), respectively. BLG QDs with p-type and n-type central doping confine holes and electrons, respectively. **(b-e)** Simulated LDOS maps for a QD with p-type and n-type central doping. The potential well and the gap size is the same in (b-c), but the interlayer potential difference polarity ($U$) is $+60$ meV for (b and d) and $-60$ meV for (c and e). The blue triangle in (b and e) and yellow triangle in (c and d) represents the orientation of the $\alpha$ and $\beta$ triangles of the BLG lattice in the TB model, respectively. Inset in (b-c) and (d-e) shows the calculated Berry curvature ($\Omega$) of valence band (VB) and conduction band (CB) near $K$ valley with $U = +60$ meV and $U = -60$ meV, respectively.

Figure 4

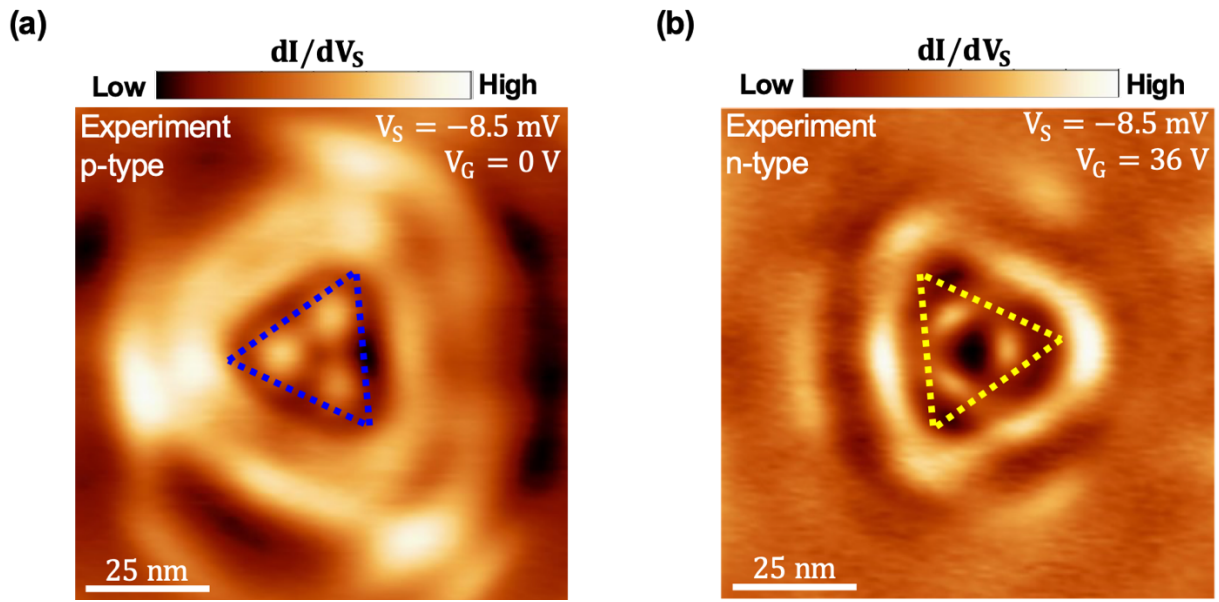

**Figure 4: $dI/dV_S$ maps reveal Berry curvature manipualtion within BLG QDs.** (**a**) Measured constant energy $dI/dV_S$ maps for a BLG QD with p-type central doping with scanning parameters $I = 0.3$ nA and $V_{ac} = 2$ mV. (**b**) Measured constant energy $dI/dV_S$ maps for a BLG QD with n-type central doping with scanning parameters $I = 0.1$ nA and $V_{ac} = 3$ mV. The blue triangle in (a) and yellow triangle in (b) represent the orientations of $\alpha$ and $\beta$ triangles of the BLG lattice, respectively.